\documentclass[12pt]{article}
\usepackage{graphicx}

\author{Christoph Best\footnote{Electronic mail: {\tt
      christoph.best@ifi.lmu.de}},\ \ Ralf Zimmer, and Joannis Apostolakis\\
        Institute for Informatics, LMU, \\
        Amalienstr. 17, 80333 M\"unchen, Germany}
\date{June 15, 2004}

\title{Probabilistic methods for predicting protein functions
       in protein-protein interaction networks}

\newcommand{\nnm}{\nonumber}

\begin{document}

\maketitle

\begin{abstract}
  We discuss probabilistic methods for predicting protein functions
  from protein-protein interaction networks.  Previous work based on
  Markov Randon Fields is extended and compared to a general
  machine-learning theoretic approach.  Using actual protein
  interaction networks for yeast from the MIPS database and GO-SLIM
  function assignments, we compare the predictions of the different
  probabilistic methods and of a standard support vector machine. It
  turns out that, with the currently available networks, the simple
  methods based on counting frequencies perform as well as the more
  sophisticated approaches.
\end{abstract}

\section{Introduction}

Large-scale comprehensive protein-protein interaction data, which have
become available recently, open the possibility of deriving new
information about proteins from their associations in the interaction
graph. In the following, we discuss and compare several probabilistic
methods for predicting protein functions from the functions of
neighboring proteins in the interaction graph.

In particular, we compare two recently published methods that are
based on Markov Random Fields \cite{Letovsky,Deng} with a prediction
based on a machine-learning appproach using maximum-likelihood
parameter estimation. It turns out that all three approaches can be
considered different versions of each other using different
approximations.  The main difference between the Markov Random Field (MRF)
and the machine-learning methods is that the former apprach
takes a global look at the network, while the latter considers each
networks node as an independent training example. However, in the
mean-field approximation required to make the MRF approach numerically
tractable, it is reduced to considering each node independently. The
local enrichment-method considered in \cite{Letovsky} can then be
interpreted as another approximation which enables us to make
predictions directly from observer frequencies, bypassing the
numerical minimization step required in the more general
machine-learning approach.

We also extend these methods by considering a non-linear
generalization for the probability distribution in the
machine-learning approach, and by taking larger neighborhoods in the
network into account. Finally, we compare the performance of these
methods to a standard Supper Vector Machine.

\section{Methods}

We consider a network specified by a graph whose nodes are proteins
and whose undirected vertices indicate interactions between the
proteins. Each node is assigned one of a set of protein functions.  In
a machine-learning approach to prediction, this assignment follows a
simple probability function depending on the protein functions in the
network neighborhood of each node and parametrized by a small set of
parameters. The learning problem is to estimate these parameters from
a given sample of assignments. The prediction can then be
performed by evaluating the probability distribution using these
parameters.

\subsection{Machine-learning approach} 
\label{sec:ml}

Assume we only consider a single protein function at a time. Node
assignments can then be chosen binary, $x\in\{0,1\}$, with $1$
indicating that a node has the function under consideration. In the
simplest case, the probability that a node $i$ has assignment $x$
depends only its immediate neighbors, and since all vertices of the
graph are equal, it can only depend on the number of neighbors $C$,
and the number of active neighbors $N$. Borrowing from statistical
mechanics, we write the probability using a potential $U(x;C,N)$
\begin{equation} \label{eq:2}
  p(x|C,N) = \frac{e^{-U(x;C,N)}}{Z(C,N)}, \qquad
  Z(C,N) = \sum_{y=0,1} e^{-U(y;C,N)}
\end{equation}
where the partition sum $Z(C,N)$ is a normalizing factor. This equation
basically expresses that the log-probabilities of $x$ are
proportional to the potential $U(x;C,N)$. In a lowest-order
approximation, we can choose a linear function for the potential:
\begin{equation} \label{eq:1}
  U(x;C,N;\alpha) = (\alpha_0 + \alpha_1 C + \alpha_2 N)x \quad.
\end{equation}
Later, we will extend this approach to more general functions.

The parameters $\alpha$ can be estimated from a set of training
samples $(x_i,C_i,N_i)$ by maximum-likelihood estimation. In this
approach, they are chosen to maximize the joint probability
\begin{equation}
  P = \prod_i p(x_i|C_i,N_i)
\end{equation}
of the training data, or equivalently, to minimize its negative logarithm
\begin{equation}
  -\log P = \sum_i \left[ \ln Z(C_i,N_i) + U(x_i;C_i,N_i) \right] \quad.
\end{equation}
Taking the partial derivative w.r.t.~to a parameter gives the equation
\begin{equation}
  -\frac{\partial P}{\partial \alpha}
  = \sum_i \left\{ - \frac{1}{Z(C_i,N_i)} \sum_{y=0,1} 
    \frac{\partial U(y,C_i,N_i)}{\partial\alpha}
    e^{-U(y,C_i,N_i)}
    + \frac{\partial U(x_i,C_i,N_i)}{\partial\alpha} \right\} \quad.
\end{equation}
The first term in the bracket is the expectation value of $\partial
U/\partial\alpha$ in the neighborhood $(C_i,N_i)$ under the
probability distributions parametrized by $(\alpha,\ldots)$:
\begin{equation}
  \left\langle \frac{\partial U(y,C_i,N_i)}{\partial\alpha}
  \right\rangle_{N_i,C_i;\alpha,\ldots} =
  \frac{1}{Z(C_i,N_i)} \sum_{y=0,1} 
    \frac{\partial U(y,C_i,N_i)}{\partial\alpha} \,
    e^{-U(y,C_i,N_i)}
\end{equation}
At the extremum, the derivative vanishes and we have the simple relation
\begin{equation}
    \sum_i \left\langle \frac{\partial U(y,C_i,N_i)}{\partial\alpha}
  \right\rangle
    = \sum_i \frac{\partial U(x_i,C_i,N_i)}{\partial\alpha} \quad.
\end{equation}
Thus, in the maximum-likelihood model, the parameters are adjusted so
that the average expectation values of the derivatives of the
potential are equal to the averages observed in the training data.
Using eq.~\ref{eq:1}, this gives the set of three equations.
\begin{eqnarray}
  \sum_i \left\{ \begin{array}{l} 1 \\ C_i \\ N_i \end{array} \right\} \,
   \langle x \rangle
 &=& \sum_i \left\{ \begin{array}{l} 1 \\ C_i \\ N_i \end{array} \right\} \, x_i 
\end{eqnarray}
where the expectation value of $x$ in the environment $(C_i,N_i)$ and
in the model parametrized by $\alpha$ is given by
\begin{equation}
  \langle x \rangle = 
  \langle x \rangle_{\alpha_0,\alpha_1,\alpha_2;C_i,N_i} 
  = \frac{e^{-(\alpha+\alpha_1 C_i+\alpha_2 N_i)}}{1+e^{-(\alpha+\alpha_1
     C_i+\alpha_2 N_i)}} \quad.
\end{equation}
Only in the simplest case, $\alpha_1 = \alpha_2 = 0$,  this equation
can be solved analytically, leading to the relation:
\begin{equation} \label{eq:3}
  \alpha = \frac{\bar x}{1-\bar x}, \qquad\mbox{with}\qquad
  \bar x = \frac{1}{n} \sum_{i=1}{n} x_i \quad.
\end{equation}
In the general case, we solve these equations numerically using a
conjugate-gradient method by explicitly minimizing the joint
probability $P$.

\subsection{Network approach}

An alternative approach to prediction starts out from considering a
given network and the protein function assignments as a whole and
assigning a score based on how well the network and the function
assignments agree. In the approach of \cite{Deng}, each link
contributes to this score with a gain $G_0$ or $G_1$, resp., if both nodes at the ends of the
link have the same function $0$ or $1$, and a penalty $P$ if they have different
function assignments. Assuming again that the log-probabilities are
proportional to the scores, this induces a probability
distribution over all joint function assignments ${\bf x}$ given by
\begin{equation}
  p({\bf x}) = \frac{1}{Z} e^{-U({\bf x})} \quad, \qquad
  Z = \sum_{\bf x} e^{-U({\bf x})}
\end{equation}
where now the normalization factor is calculated by summing over all
possible joint function assignments of the nodes.

The scoring function $U({\bf x})$ can be expressed as
\begin{eqnarray}
  U({\bf x}) &=& -\frac{G_1}{2} \sum_{i,j:(i,j)\in E}   x_i x_j
    - \frac{G_0}{2} \sum_{i,j:(i,j)\in E}   (1-x_i)\, (1- x_j)
    \\ \nnm
    &&+ \frac{P}{2} \sum_{i,j:(i,j)\in E} \left( (1-x_i) \, x_j + x_i \, (1-x_j) \right)
         + \eta_0 \sum_i x_i
    \\ \nnm
   &=& \eta_0 \sum_i  x_i
                     + \eta_1 \sum_i C_i x_i
                     + \frac{\eta_2}{2} \sum_{i,j: (i,j)\in E}  x_i x_j
\end{eqnarray}
with the parameters
\begin{equation}
  \eta_2 = - G_1 - G_0 - 2P \qquad\mbox{and}\qquad
  \eta_1 = G_0 + P \quad.
\end{equation}
In terms of statistical mechanics, this describes a ferromagnetic
system where the inverse temperature is determined by $\eta_2$ and an
external field by $\eta_0$ and $\eta_1$.  

Again, maximum-likelihood parameter estimation is performed by finding
a set of parameters $\eta = (\eta_0,\eta_1,\eta_2)$ such that the
probability of the $N$ sample configurations ${\bf x}^{(n)}$ is maximized:
\begin{equation}
  \alpha = \mathop{\rm argmax}_\alpha \sum_n^N \ln p({\bf x}^{(n)};\alpha)
  = \mathop{\rm argmin}_\alpha \left( \sum_n U({\bf x}^{(n)}) + N \ln Z(\alpha) \right)
\end{equation}
The logarithm of the partition sum appearing in the second term can
be related to the entropy by
\begin{eqnarray}
  S &=& - \sum_x p(x) \, \ln p(x)
    = \sum_{x} p(x) \, U(x) + \ln Z
  \\ \Rightarrow\qquad
    -\ln Z &=& \langle U \rangle- S = F
\end{eqnarray}
The quantity $\langle U \rangle - S$ is the thermodynamical free
energy. Maximum likelihood parameters estimation therefore corresponds
to choosing the parameters such that the energy of the given
configuration is minimized while the free energy of the system as a
whole is maximized:
\begin{equation} \label{eq:13}
  \mathop{\rm argmin}_\alpha \left( U(X;\alpha) - F(\alpha) \right) 
  = \mathop{\rm argmin}_\alpha \left( U(X;\alpha) - \langle U \rangle(\alpha) +
  S(\alpha) \right)
  \quad.
\end{equation}
Unfortunately, this requires the calculation of both the internal
energy, $\langle U \rangle(\alpha)$, and the entropy, $S(\alpha)$, of
the system and thus more or less a complete solution of the system. 

This can be avoided by employing the \emph{mean field} approximation, in
which the probability distribution $p(x)$ is replaced by a trial
distribution $p_{\rm trial}(x)$ as a product of single-variable
distributions
\begin{equation}
  p_{\rm trial}(x) = p_1(x_1) \ldots p_n(x_n)
\end{equation}
which can be completely parametrized by the expectation values $\bar
x_i$ using
\begin{equation}
  p_i(x_i) = x_i \bar x_i + (1-x_i) (1-\bar x_i)
           = \left\{ \begin{array}{ll} 1-\bar x_i & \mbox{if $x_i=0$} \\
                                   \bar x_i & \mbox{if $x_i=1$} \end{array}\right.
\end{equation}
Optimum values for the parameters $\bar x_i$ can then be estimated by
minimizing the KL entropy of $p_{\rm trial}(x)$ vs.~the true distribution
$p(x)$. 

Interestingly, this approximation removes the distinguishing feature
of the network approach, namely that the neighborhood structure (in
the sense of neghbors of neighbors) is taken into account.  The
resulting equations are very similar to the machine-learning equations
in which neighbors are treated as unrelated.

\subsection{Binomial-neighborhood approach}
\label{sec:bin}

The binomial-neighborhood approach \cite{Letovsky} is a simpler approach in which the
probability distribution $p(x|C,N)$ is chosen in such a way that it
can be directly derived from observed frequencies without the
minimization process typical for maximum-likelihood approaches. It is
based on the assumption that the distribution of active neighbors
$N_i$ of a
node $i$ follows a binomial distribution whose single probability $p$
depends on whether the node $i$ is active or not:
\begin{equation}
  p(N_i|C_i,x_i=1) = \left(\begin{array}{c} C_i \\ N_i
      \end{array} \right) \,
    p_1^{N_i} (1-p_1)^{C_i - N_i} \quad,
\end{equation}
and correspondingly for $x_i=0$ using a single probability $p_0$. This
is the assumption of \emph{local enrichment}, i.e.~that the
probability $p_1$ to find an active node around another active node is 
larger than the probability $p_0$ to find an active node around an
inactive node. Using Bayes' theorem, we can use this to calculate the
probability distribution of $x_i$:
\begin{equation}
  p(x_i|C_i,N_i) = \frac{ p(N_i|C_i,x_i) \, p(x_i|C_i) }{
                          p(N_i|C_i)} 
\end{equation}
where $p(x_i|C_i) = \bar x$ is the overall probability of observing an
active node, and
\begin{equation}
  p(N_i|C_i) = \bar x p(N_i|C_i,x_i=1) + (1-\bar x) p(N_i|C_i,x_i=0) \quad.
\end{equation}
The resulting probability distribution can be written as
\begin{equation}
  p(x_i=1|C_i,N_i) = \frac{\lambda}{1+\lambda} \qquad\mbox{and}\qquad
  p(x_i=0|C_i,N_i) = \frac{1}{1+\lambda} 
\end{equation}
with
\begin{equation}
  \lambda = \frac{\bar x}{1-\bar x} \, \frac{p_1^{N_i} \, (1-p_1)^{C_i - N_i}}{
                      p_0^{N_i} \, (1-p_0)^{C_i - N_i}}  \quad.  
\end{equation}
This can be easily rewritten in the same form as (\ref{eq:2})
\begin{equation}
  p(x_i|C_i,N_i) = \frac{1}{Z} \exp \left[-\left(  -\ln \frac{\bar x}{1-\bar x} 
   - \ln \frac{p_1}{p_0}  N_i
   + \ln \frac{1-p_0}{1-p_1} \, (C_i-N_i) \right)  \,x_i\right]
\end{equation}
The first term in the potential has the same form as (\ref{eq:3}) and
adjusts the overall number of positive sites; the two other terms
constitute a bones for having positive neighbors (proportional to $N_i$)
and a penalty for having negative neighbors (proportional to $C_i -
N_i$).

This approach evidently gives a conditional probability distribution
$p(x_i|C_i,N_i)$ of the same for as the one in the machine-learning
approach. However, the coefficient in the potential can be directly
calculated from the observed frequencies $\bar x$, $p_0$, and
$p_1$. This is only possible because we made here the assumption that
the probability distribution $p(N_i|C_i,x_i)$ is binomial. The
machine-learning approach is more flexible in that in does not have to
make this assumption and yields a true maximum-likelihood estimate
even for distributions that deviate greatly from binomial form. In
particular, the binomial distribution implies that the neighbors of a
node behave statistically independent, which might be violated in a
densely connected network, where we would expect clusters to form.

\section{Results}

To compare the different prediction methods, we chose the MIPS
protein-protein interaction database for \emph{Saccharomyces
  cerevisiae} \cite{MIPS,Uetz} and the GO-SLIM database of protein function
assignments from the Gene Ontology Consortium \cite{GO}. The latter is a
slimmed-down subset of the full gene ontology assignments comprising
32 different processes, 21 functions, and 22 cell compartments. We
focused here on the process assignments as these were expected to
correspond most closely to the interaction network. 

We compared four methods:
\begin{enumerate}
\item the binomial neighborhood enrichment from sec.~\ref{sec:bin},
\item the machine-learning maximum-likelihood method from
  sec.~\ref{sec:ml} using a linear potential (\ref{eq:1})
\item the same method with an extended non-linear potential, and
\item a standard support vector machine \cite{libsvm}.
\end{enumerate}

For the probabilistic methods, we first looked at the single-function
prediction problem in which the system is presented with a binary
assignment expressing which proteins are known to have a given
function, and then makes a prediction for an unknown protein based on
the number of neighbors that have this function.

\begin{figure}[htb]
  \begin{center}
    \includegraphics[angle=270,width=\hsize]{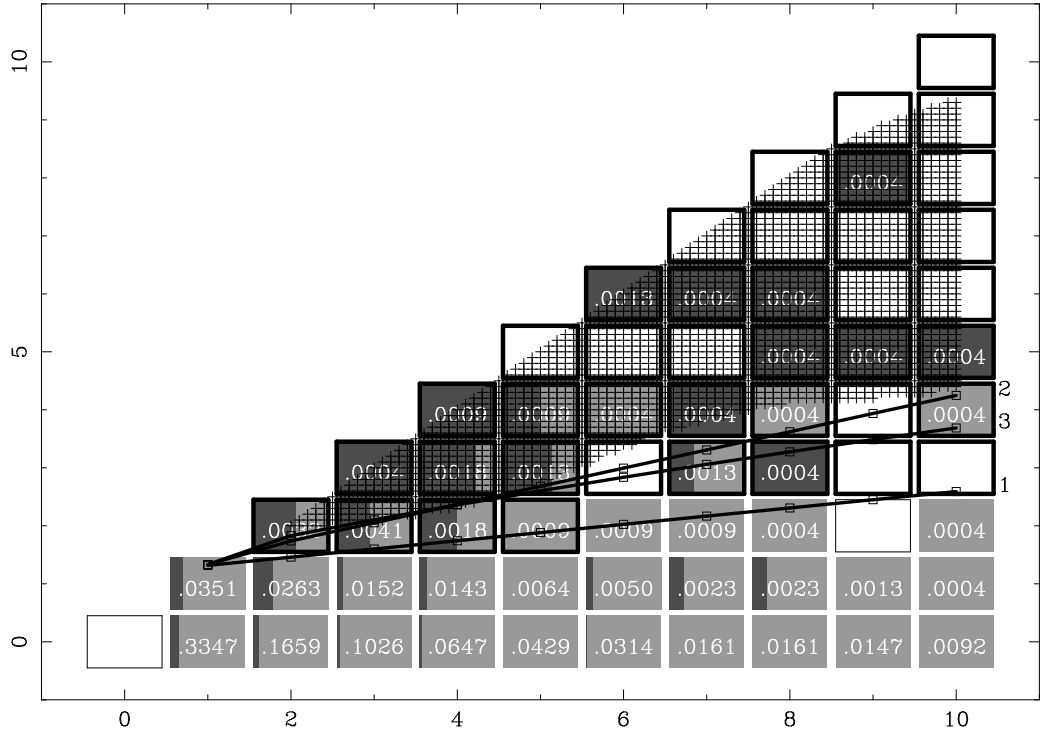}
    \caption{Glyph plot summarizing the probability distribution
      for a single-function prediction problem.
      Each box represents a possible situation of a single node,
      characterized by the total number of neighbors on the $x$-axis,
      and the number of neighbors having the funtion of interested on 
      the $y$-axis. The numbers indicate the total incidence of the
      situation, while the shading expresses how frequently the
      central node had the function of interest in that situation.
      The lines are the decision boundaries for the binomial method
      and the linear and polynomal machine-learning methods. The
      shading is the prediction region from the SVM.
    }
    \label{fig:1}
  \end{center}
\end{figure}

In this case, the local environment of a node can be described by two
numbers: $n$, the number of neighbors, and $j$, the number of
neighbors that have the function assignment under consideration. The
content of the training data set can be characterized by a glyph plot
such as in fig.~\ref{fig:1}. 

After learning the training data, the probabilistic method has
inferred a probability distribution that yields, for each pair
$(n,j)$, a probability $p(X_i=1|n,j)$ which is then utilized for
predictions. The 50\%-level of this probability, which determines the
prediction in a binary system, is indicated in fig.~\ref{fig:1} by
green lines. 

The three probabilistic predictors in fig.~\ref{fig:1} yield similar
results that differ rarely by more than one box. The main difference
is that the binomial predictor is restricted to a straight line, while
the linear and non-linear maximum-likelihood predictors can accomodate
a little turn. Linear and non-linear predictors differ only minimally.

\begin{figure}[htb]
  \begin{center}
    \includegraphics[width=\hsize]{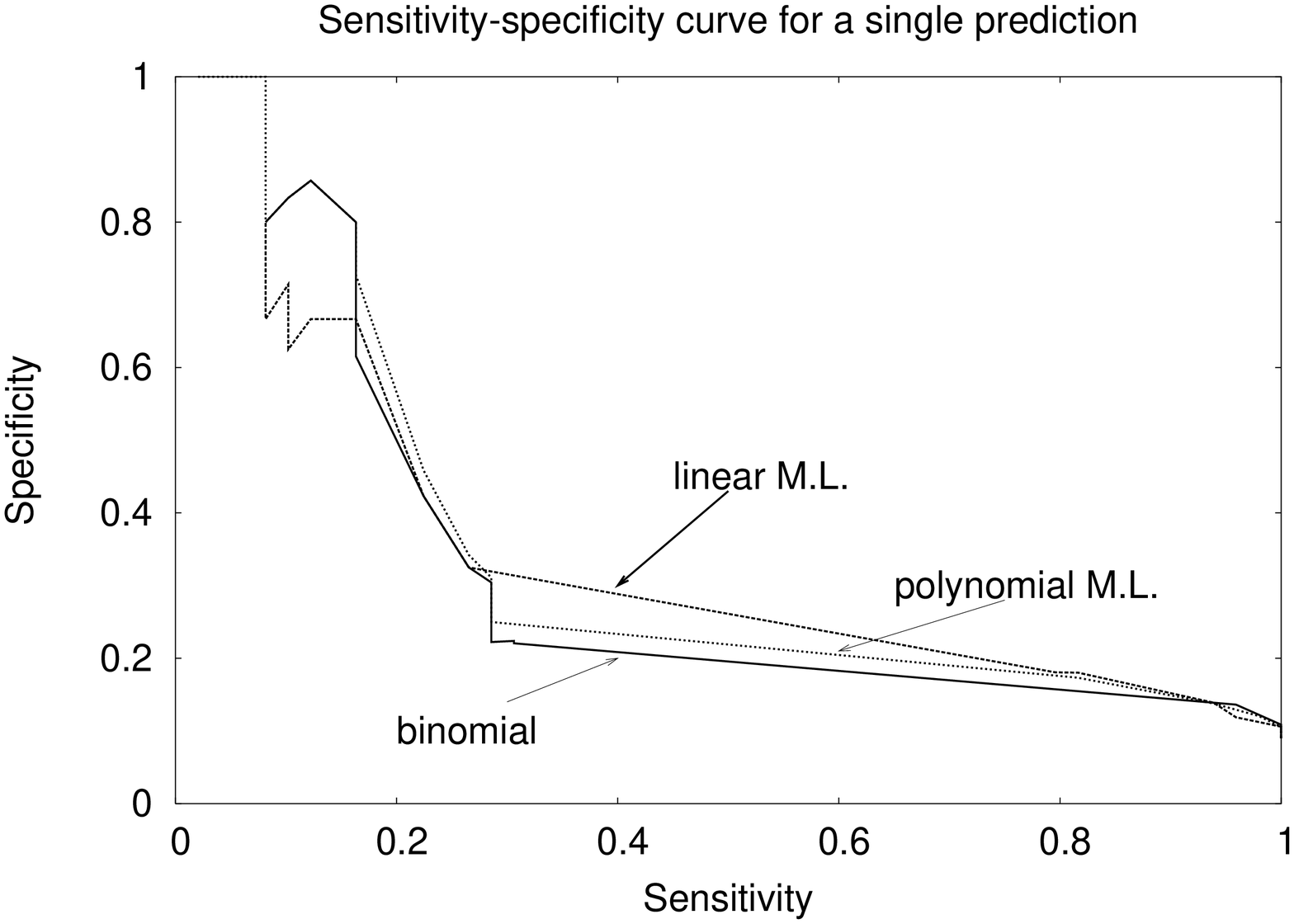}
    \caption{Sensitivity-specificity curve for the three probabilistic
    prediction methods for a single-function prediction.}
  \end{center}
    \label{fig:2}
\end{figure}

Finally the prediction from a support vector machine that was trained
on the same single-function data set is indicated by a shaded area
marking all those $(n,j)$ for which the SVM returned a positive
prediction. The border of this area very closely follows the linear
and non-linear M.L.~predictors.

Fig.~\ref{fig:2} shows a sensitivity-specificity curve using five-fold
cross validation for single-function prediction using the
probabilistic predictors. Again, all three curves follow each other
quite closely, with a slight edge for the nonlinear M.L.~predictor.

The preceding discussion applied to the problem of single function
prediction. To perform full prediction, we generated each of the three
predictors separately for each function and chose, for each protein
with an unknown function, the prediction with the largest probability.
For simplicity, this approach does not take into account possible
correlations between different protein functions. However, such
correlations were taken into account for the support vector machine,
which generated a full set of cross-predictors (predicting function
$i$ with neighbors of type $j$). 

\begin{figure}[htb]
  \begin{center}
    \includegraphics[width=\hsize]{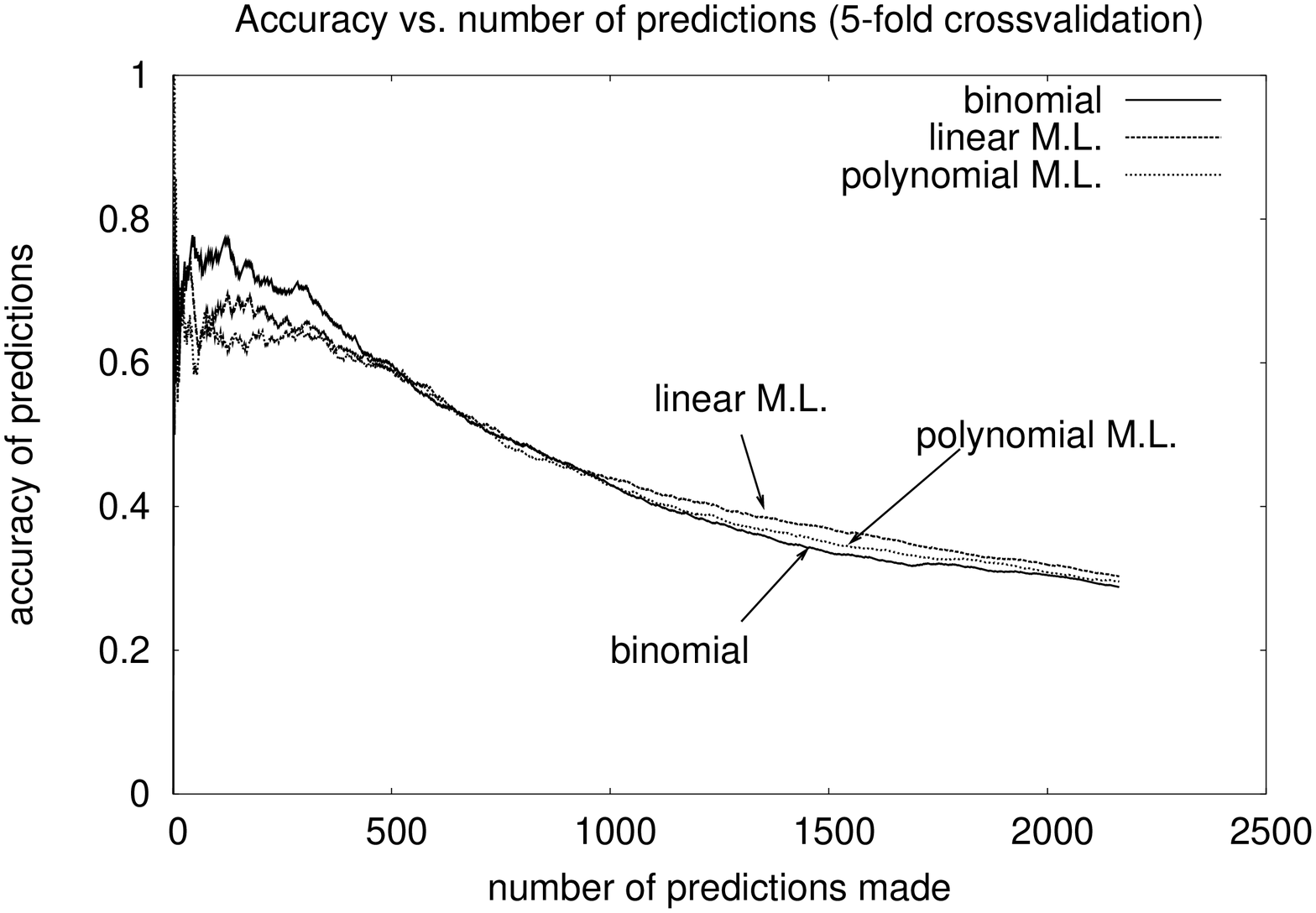}
    \caption{Accuracy of multiple-function prediction as a function of
      the number of predictions made using the three probabilistic prediction methods.}
  \end{center}
    \label{fig:3}
\end{figure}

In the probabilistic case, each predictor does not only provide us
with a yes-no decision, but also with a probability for the
prediction. We can use the information to restrict the predictions to
highly probable ones.  Fig.~\ref{fig:3} shows the accuracy of the
prediction as a function of how many predictions are made with
different cut-offs in the predicted probability. Again, all three
curves closely follow each other, with maybe a small but unsignificant
edge of the linear M.L.~predictor. The predictions from all predictors
including the SVM were similar, and combining them would not have
improved predictive accuracy.

\begin{table}[htb]
  \centering
  \begin{tabular}{|l|l|l|}
    \hline
    METHOD & \#SUCCESS & accuracy \\
    \hline
    binomial classifier & 623 & 31\% \\
    linear M.L.\ classifier & 655 & 33\% \\
    nonlinear M.L.\ classifier & 640 & 31.7\% \\
    linear SVM classifier & 601 & 29.8\% \\
    \hline
    randomized network & 101 & 11.4\% \\
    \hline
    \hline
    binomial classifier, process &  & 32.5\%\\
    randomized network & & 8.7\% \\
    \hline
  \end{tabular}
  \caption{Prediction accuracy in five-fold cross validation for the
    yeast data set.}
  \label{tab:1}
\end{table}

Finally, the success rates for all predictors are shown in table
\ref{tab:1} using five-fold cross-validation on a data set of 2014
unique function assignments for the yeast proteome. It turns out that
all four methods perform closely, with success rates between 30 and
33\%. This compares to the null-hypothesis of prediction in a
randomized network, in which we would have a success rate of 11\% for
these data. The protein-protein interaction data therefore roughly
triples the prediction success over a random network. However, all
methods, from the simple, counting-based binomial classifier to the
full support vector machine, perform similarly.

We also extended our methods to take larger neighborhoods (second and
higher-order neighbors) into account, but failed to substantially
improve predictive power.

Finally, we also performed protein function prediction on a recently
published protein-interaction network for {\em Drosophila
  melanogaster} \cite{Droso}, with similar results. 

\section{Discussion}

We compared different probabilistic approaches to predicting protein
functions in protein interaction networks. Under closer analysis, the
different Markov Random Field methods in the literature can be related
to a basic machine-learning approach with maximum-likelihood parameter
estimation. Using real data, they exhibit similar performance, with
simple methods performing as well as more complex ones. This might
indicate limits on the functional information contained in
protein-protein interaction networks.

A standard support vector machine gave similar result, though it was
equipped with more information, namely the frequencies of all function
classes in the neighborhood. The additional information did neither improve nor
harm predictive performance.

\end{document}